# DIGITAL EPIDEMIOLOGY: A REVIEW

David Pastor-Escuredo[1,2,3,*]

[1]LifeD Lab, Spain
[2]UCL, UK
[3]UNICEF
*corresponding author: david@lifedlab.org

**ABSRACT**

The epidemiology has recently witnessed great advances based on computational models. Its scope and impact are getting wider thanks to the new data sources feeding analytical frameworks and models. Besides traditional variables considered in epidemiology, large-scale social patterns can be now integrated in real time with multi-source data bridging the gap between different scales. In a hyper-connected world, models and analysis of interactions and social behaviors are key to understand and stop outbreaks. Big Data along with apps are enabling for validating and refining models with real world data at scale, as well as new applications and frameworks to map and track diseases in real time or optimize the necessary resources and interventions such as testing and vaccination strategies. Digital epidemiology is positioning as a discipline necessary to control epidemics and implement actionable protocols and policies. In this review we address the research areas configuring current digital epidemiology: transmission and propagation models and descriptions based on human networks and contact tracing, mobility analysis and spatio-temporal propagation of infectious diseases and *infodemics* that comprises the study of information and knowledge propagation. Digital epidemiology has the potential to create new operational mechanisms for prevention and mitigation, monitoring of the evolution of epidemics, assessing their impact and evaluating the pharmaceutical and non-pharmaceutical measures to fight the outbreaks. Epidemics have to be approached from the lens of complexity as they require systemic solutions. Opportunities and challenges to tackle epidemics more effectively and with a human-centered vision are here discussed.

## 1. INTRODUCTION

Epidemiology is the field that encompasses the study of the distribution, prevalence and etiology of human diseases. (Green, Freedman, & Gordis, 2000; Salathe et al., 2012). Although data and models have been always part of epidemiology (Anderson, Anderson, & May, 1992), the appereace of new sources of Big Data and technology (Bell, Hey, & Szalay, 2009) have enabled computational frameworks and opportunities to increase impact and knowledge (Salathe et al., 2012). In parallel to the appearance and use of new data sources, the growth of Artificial Intelligence and specially Machine Learning techniques (Bullock, Pham, Lam, & Luengo-Oroz, 2020; LeCun, Bengio, & Hinton, 2015), are giving rise to multiple methodologies and applications that can be categorized as the emergent digital epidemiology.

The studies of epidemiology have been grounded in data collected in clinical practice and field work (Hernán & Robins, 2006). First steps in digital epidemiology were posible thanks to data on the Internet, specially web searches (e.g. Google flue trends) from individuals (Brownstein, Freifeld, & Madoff, 2009; Cervellin, Comelli, & Lippi, 2017; Eysenbach, 2009; Ginsberg et al., 2009). This trend generated the first difficulties in methodologies and epistemiology of the new field (Cook, Conrad, Fowlkes, & Mohebbi, 2011; D. Lazer, Kennedy, King, & Vespignani, 2014; Olson, Konty, Paladini, Viboud, & Simonsen, 2013). Currently, we count on many sources such as social media, social networks, mobile apps and other services that generate data (Salathé, 2018). This progress in the digitalization is getting more prone because of the COVID-19 pandemic caused by SARS-CoV-2, accelerating the adoption of the digital tech in all sectors that have had a slower transition to the digital.

For this reason, digital epidemiology is in charge of understanding the dynamics of patterns, both social and clinical, of affected people by the disease and the causes of these patterns (Salathé, 2018). According to the definition of WHO, epidemiology is the study of the distribution and the determinants of the estates and events related with health and the application to disease control and management and other health challenges. Therefore, epidemiology has a pragmatic dimension aimed to improve response systems against epidemics including prevention, management, mitigation and preparation to future epidemics and waves. Besides, epidemiology, due to its complexity and importance is promoting other theories and techniques. Digital epidemiology is not only about new technology, but mostly about the scope of epidemiology to manage complexity of diseases and their factors: biological, social and environmental. Scope is larger because more data is used and analyzed including data that was not thought or designed for health applications (Salathé, 2018). In this document, we overview the work areas and the ongoing work along with the most important contributions where COVID-19 has been a disruption point (Vespignani et al., 2020).

## 2. MODELS AND NETWORKS

There are two types of epidemiological models: models based on equations and models based on agents (ABMs). Models based on equations assume homogeneity and similar collective behavior (Anderson et al., 1992). Progress on computation has enabled ABMs that can model heterogeneity in epidemics (Aleta et al., 2020; Espana et al., 2020). Both types of models are based on the conceptualization of the disease through different states, being SIR (Susceptible, Infected, Recovered) the most used and extended. There are several extensions to introduce new complexities and details. For instance, the model SEIR inserts the state Exposed that comprises people infected in incubation process. Each state can be parametrized towards quantifying the transition between states given biological and social criteria inferred from clinical data, surveys and questionnaires. Typically, models output different variables including the forecast of infected cases, forecast of deaths and epidemiological parameters such as the reproductive number $R_0$. However, inputs may greatly differ between models. The reproductive number $R_0$ is the average number of contagions generated by each person and R is the full distribution of ($R_n$) in each node. Models integrate diseases characteristics, temporality and volume of the epidemic.

A recent key element in sophisticated epidemiological models is to introduce the complexity of contacts between people (contact matrix) as multi-layer networks, so the disease depends on the structure, properties and topology of the network. Even when all scales are interlinked (from biochemical to social), epidemiology based on networks is useful to predict the propagation of the disease and implement policies. Networks allow modelling the behavioral component of the disease through the network itself and its dynamics: percolation and diffusion.

Considering the topology of the network, a disease can propagate with different velocities and scopes (de Arruda, Petri, Rodrigues, & Moreno, 2020; Moreno, Pastor-Satorras, & Vespignani, 2002; Pastor-Satorras & Vespignani, 2001b). Several studies have proposed "scale-free" networks, the distribution of the number edges of each node follows and exponential law, which represents reasonably a real-world scenario of how people is interconnected. The epidemiological studies based on this model shows the absence of an epidemiological threshold and large heterogeneity in the behavior of the network (Pastor-Satorras & Vespignani, 2001b). This is due to the diffusion in scale-free networks that changes $R_n$ of each node in time. The epidemiological threshold is defined as the number of cases where $R_0$ evolves as an exponential curve. However, it is unclear why scale-free networks do not have epidemiological threshold. Models that are more realistic and consider constraints in the size of "scale-free" have a threshold and still heterogeneous behavior (Pastor-Satorras & Vespignani, 2001a, 2002).

Percolation is understood as the process from which nodes of the network transition to another state. The structure is key for this dynamic phenomenon due to the compartments of the network (Meyers, Newman, & Pourbohloul, 2006; Mark EJ Newman, 2002). Through percolation is possible to model the dynamics of disease spreading and when the epidemiological threshold is passed. Diffusion is the other dynamic phenomenon that can be studied. Diffusion is the process from which several nodes are reached from one node and depends on the topology of the network and the dynamics of the disease. Predicting diffusion is critical to slow down and stop epidemics.

Networks are part of epidemiological models by using contact tracing matrices that enable the reconstruction of the contagion network if they are properly designed and collected. These matrices can be stratified (different age groups and gender groups) and multi-layer (if several subpopulation are labelled, i.e. work travels, home travels, leisure travels, etc) (Aleta et al., 2020). Recent methodological advances comprise the use of hyper-graphs with links that connect several nodes instead of pair-to-pair links (de Arruda, Petri, & Moreno, 2020).

Thus, networks are used to build risk forecast and propagation forecast systems, including the analysis of transportation hubs (Colizza, Barrat, Barthelemy, Valleron, & Vespignani, 2007), confirming that heterogeneity favors the propagation as it is easier to percolate. Consequently, actions to prevent, stop and contain epidemics have to reduce and make more homogeneous the degree of the nodes in all scales for a given time window, so the epidemic is easier to control (Chinazzi et al., 2020; Martín-Calvo, Aleta, Pentland, Moreno, & Moro, 2020; Zhang et al., 2020).

Network analysis has witnessed a new revolution during COVID-19 due to the new data sources acquired via Bluetooth and geo-location of mobile devices that enable GPS-based and proximity-based contact tracing to obtain dynamic and high-resolution matrices (Ferretti et al., 2020). Multi-source networks will enable multi-partita networks where interactions between people and locations can be represented.

Networks are also useful to track and understand recovery and resilience, which in this case is favored by heterogeneity processes of recovery within the network (Boy, Pastor-Escuredo, Macguire, Jimenez, & Luengo-Oroz, 2019; de Arruda, Petri, Rodrigues, et al., 2020). Another application is to understand the interaction among concurrent (Sanz, Xia, Meloni, & Moreno, 2014).

## 3. MOBILITY AND PROPAGATION

Mobility has a direct impact on disease propagation air-borne or vector-borne. The mobility studies have a long tradition but have been hampered by the lack of dynamic and fine-grained data to differentiate types of mobility (Prothero, 1977). Last decades we have witnessed different types of mobility: tourism, events, business, long-term labor or students' mobility. It is not longer possible to study different layers of mobility through surveys and static data. The analysis of mobility and spatial characteristics of diseases depends on the availability and resolution of longitudinal data (Eubank et al., 2004; Gonzalez, Hidalgo, & Barabasi, 2008; Riley, 2007).

Human mobility is multi-scale in temporal and spatial dimensions (Candia et al., 2008; Gonzalez et al., 2008; Pastor-Escuredo & Frias-Martinez, 2020; Zufiria et al., 2018). Human mobility is also multilayered depending on the population flows. These layers are interconnected and each of them are propagated through a "social medium" (Balcan, Colizza, et al., 2009). The structure of mobility has an amplifier effect in the propagation due to diffusion and percolation if it is not properly managed. First studies in mobility as epidemiological factor were focused on the global scale based on demographic data and international mobility statistics (Balcan et al., 2010). The temporal resolution of data in these studies only allowed studying seasonal variability (Balcan, Hu, et al., 2009). The models used were gravitation-driven model and radiation-driven model (Simini, González, Maritan, & Barabási, 2012). However, these models only work under strong assumptions and it is difficult to make them work in epidemiological practice (Perrota, 2018).

In vector-borne diseases, such as malaria or dengue, or diseases transmitted through air and water, small scale mobility affects to the exposition of people to the disease whereas large-scale mobility affects to the introduction, reinsertion and circulation of the contagions and even the global propagation (Buscarino, Fortuna, Frasca, & Latora, 2008; Lynch & Roper, 2011; Stoddard et al., 2009). Frequently, diseases induce a systemic change in mobility with hard-to-control impact (Meloni et al., 2011). For this reason, it has been identified the need to create monitoring mechanisms of mobility based on high-resolution mobile devices data.

Mobile phone data are generated from telecom operators and contain geolocation of calls and connections to the Internet. Also, geolocation services for smartphones allow capturing mobility traces (i.e. Cuebiq, Foursquare, etc). The temporal resolution of this data is very high (Blondel, Decuyper, & Krings, 2015; Pulse, 2012). The data requires an anonymization and aggregation process to preserve privacy (De Montjoye, Hidalgo, Verleysen, & Blondel, 2013; Pulse, 2015).

High-resolution longitudinal allowed the characterization "hotspots" and optimize the location of actions to prevent and stop the diseases (Bejon et al., 2010; Dolgin, 2010). However, mobile phone data has enabled the revolution of mobility for epidemics (Bengtsson et al., 2015; Bengtsson, Lu, Thorson, Garfield, & Von Schreeb, 2011; Tatem et al., 2014; Tizzoni et al., 2014; Wesolowski et al., 2012; Wesolowski et al., 2015). Sources and sinks characterization was the first and one the most notable to decipher the structure of propagation and generate risk maps separate from vector density maps (Wesolowski et al., 2012). However, a more detailed study based on mobility flows descriptors (Pastor-Escuredo & Frias-Martinez, 2020), can help understand the dynamics of risk and super-propagation phenomena.

Among the mobility phenomena, cultural events in many regions of the world have been analyzed as high-impact events in epidemics (Finger et al., 2016). Long-term mobility analysis and mobility profiles are useful tools to understand the dynamics of the epidemics in a disaggregated way (UNICEF, 2020). Disaggregation is necessary to understand the socio-economics of the epidemics (Martín-Calvo et al., 2020) and the relationships with other sectors such as work, tourism (Belderok, Rimmelzwaan, Van Den Hoek, & Sonder, 2013; Chinazzi et al., 2020) or agriculture and the rural-urban migration (Zufiria et al., 2018). Finally, mobile phone data and survey data can be integrated to have high spatio-temporal and demographic resolutions (Wesolowski et al., 2014).

During SARS-CoV-2, the number of applications and use cases of mobile phone data has increased (Oliver et al., 2020), including distancing and lockdown measures: lockdown enforcement, measuring the epidemiological impact of the lockdowns and distancing and evaluating the measures for re-opening (Martín-Calvo et al., 2020).

## 4. INFODEMICS

Decision-making during pandemics is key for good response and management and stop negative effects. The asymmetry of negative impact requires additional actions to avoid systemic risk (Norman, Bar-Yam, & Taleb, 2020; Taleb, 2019; Taleb, Bar-Yam, Douady, Norman, & Read, 2014). Decision-making demands the right information with the right timing (Greenwood, Howarth, Escudero Poole, Raymond, & Scarnecchia, 2017), for this reason, information propagation during pandemics has become a key use for United Nations General Secretary (General, 2020). SARS-CoV-2 has been marked by the spread of fake news and news that generate division (Bakir & McStay, 2018; D. M. Lazer et al., 2018; Shu, Sliva, Wang, Tang, & Liu, 2017). When this situation gets more severe in moments of crisis, it becomes an infodemic (Hua & Shaw, 2020; Vaezi & Javanmard, 2020; Zarocostas, 2020).

Several works have been done to study information propagation, specially rumor and fake news, using analogies of disease spreading across complex networks (Miritello, Moro, & Lara, 2011; Morales, Borondo, Losada, & Benito, 2014; Nekovee, Moreno, Bianconi, & Marsili, 2007). For instance, through network analysis it is possible to discover who are the leaders of the social media and their influence in information propagation (Bodendorf & Kaiser, 2009; Mark Ed Newman, Barabási, & Watts, 2006; Pastor-Escuredo & Tarazona, 2020) and quantify viral processes and info spreaders (Borge-Holthoefer, Meloni, Gonçalves, & Moreno, 2013). Beyond information propagation, semantics analysis is a useful tool to classify text. New Deep Learning tools (LeCun et al., 2015) are making this task accurate and scalable (Popat,

Mukherjee, Yates, & Weikum, 2018; Ruchansky, Seo, & Liu, 2017; Singhania, Fernandez, & Rao, 2017).

Facing risk of hatred content, it is necessary to highlight the need of propagation positive information, being constructive through the social media and the networks. Information empowers the population to take better decisions at the individual and the collective levels. Information can help people keep their environment safe. Furthermore, it helps building up resilience and increase socio-economic impact driven by Collective Intelligence (Luengo-Oroz et al., 2020).

Information gathered by citizens helps manage risk and understand the epidemic better (Leung & Leung, 2020; Sun, Chen, & Viboud, 2020), feeding computational systems and models that deal with probabilities beyond demographic and clinical data. In this sense, new sensors to monitor variables and dynamic changes in the population are necessary. There exist already several tools to classify disease analyzing coughing (Imran et al., 2020) or problems with the smelling (Menni et al., 2020). Finally, new channels between authorities and population are necessary to build up trust and improve response.

## 5. ARTIFICIAL INTELLIGENCE-DRIVEN POLICY

### 5.1. Prediction and prevention

An early and rapid response minimizes and mitigates the impact f the epidemics. Models are principally used to predict the evolution of the epidemics. The prediction is based on the area of influence, the temporal curve and R. The models are expressed in terms of variables like population density, age and gender, implying limitations in the understanding of the epidemics to predict propagation and impact. Variables like vulnerability, socio-economic inequality or WASH infrastructure are key to have more effective models. Epidemiology complexity rises due to the variability and the complexity of the ecosystems where the epidemics propagate hampering the use of local models. Some efforts have been made to develop clustering strategies to have epidemiological profiles at the geographical model (Carrillo-Larco & Castillo-Cara, 2020; Hartono, 2020; Hu, Ge, Jin, & Xiong, 2020). Furthermore, models need to be more dynamic to change policy locally and internationally. This implies better data infrastructure (health data spaces) and better models to move towards deployment and production phases Furthermore, process and model parameters backtracking is necessary to perform causality analysis of test clinical evidence.

Apart from modelling, researchers have applied Machine Learning tools to predict the behavior of epidemiological curves (D. Liu et al., 2020; Zou & Hastie, 2005) and analyze temporal patterns (Bandyopadhyay & Dutta, 2020; Huang, Chen, Ma, & Kuo, 2020). These frameworks have potential for policy making during distancing, lockdown and reopening. Tot rain these models it is necessary data that is not often available (Fong, Li, Dey, Crespo, & Herrera-Viedma, 2020). For instance, during the crisis of SARS-CoV-2, researchers have used data from other disease like flu, even when the behavior is very different (Lu, Hattab, Clemente, Biggerstaff, & Santillana, 2019; Yang, Santillana, & Kou, 2015). In other cases, it is necessary to use small data (Bullock et al., 2020).

Some diseases have a strong environmental component, for instance, increasing the density of vectors (Rajarethinam et al., 2019; Wesolowski et al., 2014). Thus, it is important to integrate environmental data and social models with high resolution and real-time. For diseases where the transmission is mainly from individual-to-individual, it is necessary to model asymptomatic cases and their contribution to the propagation (Mizumoto, Kagaya, Zarebski, & Chowell, 2020). There is new research to identify biomarkers and have clinical studies to control asymptomatic cases and understand different immunity, being necessary to configure model parameters (Kermali, Khalsa, Pillai, Ismail, & Harky, 2020; Nicholas et al., 2020; Shi et al., 2020).

Super-propagation has become also central because it has been observed a great variability in the distribution of R (dispersion k), giving rise to super-propagation events and spots. The role of super-propagators at the individual level is also important and can only be studied through contact-tracing matrices.

The society needs new tools to manage systemic risk of epidemics. This implies managing better the information taking advantage of social systems exploiting complexity. Risk is multi-dimensional and even though the health response is the most important phase, it is necessary to leverage economic risk, social inequalities, drawbacks with rights and freedom and the effects on cognition and psychological state of the population.

### 5.2. Impact tracking and assessment

Non-pharmaceutical measures have become very relevant including lockdowns, distancing, contact tracing and mobility analysis (Eames & Keeling, 2003). The objective of these measures is so reduce $R_0$ (average of distribution R) being the output of predictive models (Dietz, 1993; Q.-H. Liu et al., 2018). Super-propagation is an additional challenge that has not yet been included in epidemiological models.

The strategies of distancing are restrictive depending on the transmission medium, the morbidity and the mortality of the disease (Glass, Glass, Beyeler, & Min, 2006). To track th effect of these strategies now we can use other data such as mobile phone data (Oliver et al., 2020), Internet searches and social media (Lampos et al., 2020) and other data generated by mobile devices. However, this kind of systems are not fully implemented in our world. Some agencies are developing workflows to assess the impact on the most vulnerable populations and have a global understanding of the social dimension of epidemics (UNICEF, 2020).

One of the new systems is the digital contact tracing based on Bluetooth apps or physical proximity derived from GPS location. There are several architectures, centralized (PEPP-PT) and decentralized (DP-3T) to manage the info about risk of contagion. These techniques allow, given privacy and security mechanisms, generating suitable contact matrices with high disaggregation.

Information curation is another key process to avoid negative effects (WHO, 2020). Ad-hoc systems are normally better than general digital platforms to ensure a responsible flow of information and data (Mejova & Kalimeri, 2020; Singh et al., 2020). Dedicated chatbots and curation pipelines are part of the new epidemiology (Bullock et al., 2020).

### 5.3. Evaluation of epidemiological measures

Evaluation of measures is key to improve response systems for the short and long term. Evaluation is key for governance and policy, so the mechanisms have to be trustful and transparent to measure the impacts. COVID-19 pandemic has speed up the innovation in this area due to the severity of the distancing and lockdowns and their socio-economic impact. Deep

Learning algorithms have been used to simulate scenarios of the pandemic at a global scale (Hu, Ge, Li, et al., 2020) and measure the effects of lockdowns (Dandekar & Barbastathis, 2020). Mobile phone data has been used to measure the effects of the measures in geographical áreas but also in meta-populations and population target groups (Martín-Calvo et al., 2020; UNICEF, 2020). A global challenge is to isolate the effect of each mitigation measure to be quantified and evaluated with data (Cowling et al., 2020; Davies et al., 2020; Flaxman et al., 2020; Lai et al., 2020). Global pandemic has shown the need to make dynamic policy and update the measures using computational models in nearly real-time (Aleta et al., 2020).

## 6. CONCLUSION

Epidemics are multi-dimensional: molecular, clinical, social and political. Social dimension modulates the propagation and impact of the epidemics and pandemics. A better understanding of social systems is necessary to create new mechanisms based on collective action and efforts to avoid future pandemics. A better sustainable development will make us more resilient, robust and anti-fragile to face diseases. It has been clear that inequality is an important factor that accelerates disease propagation and their impact making the system more fragile and implying latent systemic risk (Norman et al., 2020; Taleb, Read, Douady, Norman, & Bar-Yam, 2014). It is still unclear if human conditions are variable enough to present a problem for global immunity. Super-propagation phenomena has been also shown as a systemic problem for pandemics and we do not have the tools to tackle them (Kupferschmidt, 2020; Lloyd-Smith, Schreiber, Kopp, & Getz, 2005). Epidemiological policy must act at different levels, from local aid to global governance mechanisms (Mello & Wang, 2020). We still need to progress of complex multi-scale systems for acquiring data, diagnostics and delivering aid in real-time.

Assessing measures to stop epidemics present still several epistemological, operational and political challenges. For instance, we should think in experimentation and simulations and isolate factors (pharmacological, non-pharmacological, social, political, economic, etc) to quantify how each measure contributes and also assess the synergies of integrated policy (Flaxman et al., 2020; Lai et al., 2020). Data collection systems should be started designed and implemented as part of the protocols for pandemics. Data integration is still necessary to deploy more descriptive and accurate models (Wesolowski, Buckee, Engø-Monsen, & Metcalf, 2016). SARS-CoV-2 pandemic has been a partial revolution (Ferretti et al., 2020). Next challenge is a better international system to control epidemics which implies not only regulatory issues but more technology to be prepared to stop down future epidemics (Kuhn, Beck, & Strufe, 2020; Mello & Wang, 2020; Vinuesa, Theodorou, Battaglini, & Dignum, 2020). Cross-disciplinary research is also key for better policies (Luengo-Oroz et al., 2020), including crossing molecular and clinical research with technological innovation (Bullock et al., 2020).

Communication is still failing as no new channels have been implemented. Population has been reluctant to apps because they have not been built within a trust-based ecosystem. Innovation in communication channels and novel dashboards is an important area of research (Dong, Du, & Gardner, 2020). Artificial Intelligence has to exploit the upcoming 5th Industrial Revolution to design better systems for response and decision making in all layers of the society including policy makers and citizens (Luengo-Oroz et al., 2020) and deliver Collective Intelligence platforms to empower people and amplify collective efforts (Aleks Berditchevskaia, 2020).

Ethical issues have arisen because of the exhaustive use of technology in some countries during the SARS-CoV-2 pandemic (Berman, Carter, Herranz, & Sekara, 2020; Pastor-Escuredo, 2020; Vinuesa et al., 2020). Privacy and rights have suffered an important debate because there was not sufficient reasoning on these topics (Mello & Wang, 2020). Ethical frameworks based on principles are necessary to leverage technologies for sustainable development and emergencies (Pastor-Escuredo, 2020). Public-private partnerships have not been fully leveraged for response although some data sharing projects have been implemented. This means that the Data Revolution has not been fully effective yet to fight epidemics as it was expected a decade ago. After SARS-CoV-2 we will probably witness several technological and social revolutions including the future of work (Malone, 2004).

We still must work on implementing epidemiological policies, technology and mechanisms to fight future epidemics and progress towards a more sustainable and resilient society (General, 2019).

## REFERENCES

Aleks Berditchevskaia, K. P. (2020). Coronavirus: seven ways collective intelligence is tackling the pandemic. Retrieved from https://theconversation.com/coronavirus-seven-ways-collective-intelligence-is-tackling-the-pandemic-133553

Aleta, A., Martín-Corral, D., y Piontti, A. P., Ajelli, M., Litvinova, M., Chinazzi, M., . . . Merler, S. (2020). Modelling the impact of testing, contact tracing and household quarantine on second waves of COVID-19. *Nature Human Behaviour*, 1-8.

Anderson, R. M., Anderson, B., & May, R. M. (1992). *Infectious diseases of humans: dynamics and control*: Oxford university press.

Bakir, V., & McStay, A. (2018). Fake news and the economy of emotions: Problems, causes, solutions. *Digital journalism, 6*(2), 154-175.

Balcan, D., Colizza, V., Gonçalves, B., Hu, H., Ramasco, J. J., & Vespignani, A. (2009). Multiscale mobility networks and the spatial spreading of infectious diseases. *Proceedings of the National Academy of Sciences, 106*(51), 21484-21489.

Balcan, D., Gonçalves, B., Hu, H., Ramasco, J. J., Colizza, V., & Vespignani, A. (2010). Modeling the spatial spread of infectious diseases: The GLobal Epidemic and Mobility computational model. *Journal of computational science, 1*(3), 132-145.

Balcan, D., Hu, H., Goncalves, B., Bajardi, P., Poletto, C., Ramasco, J. J., . . . Van den Broeck, W. (2009). Seasonal transmission potential and activity peaks of the new influenza A (H1N1): a Monte Carlo likelihood analysis based on human mobility. *BMC medicine, 7*(1), 45.

Bandyopadhyay, S. K., & Dutta, S. (2020). Machine learning approach for confirmation of covid-19 cases: Positive, negative, death and release. *medRxiv*.

Bejon, P., Williams, T. N., Liljander, A., Noor, A. M., Wambua, J., Ogada, E., . . . Färnert, A. (2010). Stable and unstable malaria hotspots in longitudinal cohort studies in Kenya. *PLoS medicine, 7*(7).

Belderok, S.-M., Rimmelzwaan, G. F., Van Den Hoek, A., & Sonder, G. J. (2013). Effect of travel on influenza epidemiology. *Emerging infectious diseases, 19*(6), 925.

Bell, G., Hey, T., & Szalay, A. (2009). Beyond the data deluge. *Science, 323*(5919), 1297-1298.

Bengtsson, L., Gaudart, J., Lu, X., Moore, S., Wetter, E., Sallah, K., . . . Piarroux, R. (2015). Using mobile phone data to predict the spatial spread of cholera. *Scientific reports, 5*, 8923.

Bengtsson, L., Lu, X., Thorson, A., Garfield, R., & Von Schreeb, J. (2011). Improved response to disasters and outbreaks by tracking population movements with mobile phone network data: a post-earthquake geospatial study in Haiti. *PLoS medicine, 8*(8).

Berman, G., Carter, K., Herranz, M. G., & Sekara, V. (2020). *Digital contact tracing and surveillance during COVID-19. General and child-specific ethical issues*. Retrieved from

Blondel, V. D., Decuyper, A., & Krings, G. (2015). A survey of results on mobile phone datasets analysis. *EPJ Data Science, 4*(1), 10.

Bodendorf, F., & Kaiser, C. (2009). *Detecting opinion leaders and trends in online social networks.* Paper presented at the Proceedings of the 2nd ACM workshop on Social web search and mining.

Borge-Holthoefer, J., Meloni, S., Gonçalves, B., & Moreno, Y. (2013). Emergence of influential spreaders in modified rumor models. *Journal of Statistical Physics, 151*(1-2), 383-393.

Boy, J., Pastor-Escuredo, D., Macguire, D., Jimenez, R. M., & Luengo-Oroz, M. (2019). Towards an understanding of refugee segregation, isolation, homophily and ultimately integration in Turkey using call detail records. In *Guide to Mobile Data Analytics in Refugee Scenarios* (pp. 141-164): Springer.

Brownstein, J. S., Freifeld, C. C., & Madoff, L. C. (2009). Digital disease detection—harnessing the Web for public health surveillance. *The New England journal of medicine, 360*(21), 2153.

Bullock, J., Pham, K. H., Lam, C. S. N., & Luengo-Oroz, M. (2020). Mapping the landscape of artificial intelligence applications against COVID-19. *arXiv preprint arXiv:2003.11336*.

Buscarino, A., Fortuna, L., Frasca, M., & Latora, V. (2008). Disease spreading in populations of moving agents. *EPL (Europhysics Letters), 82*(3), 38002.

Candia, J., González, M. C., Wang, P., Schoenharl, T., Madey, G., & Barabási, A.-L. (2008). Uncovering individual and collective human dynamics from mobile phone records. *Journal of physics A: mathematical and theoretical, 41*(22), 224015.

Carrillo-Larco, R. M., & Castillo-Cara, M. (2020). Using country-level variables to classify countries according to the number of confirmed COVID-19 cases: An unsupervised machine learning approach. *Wellcome Open Research, 5*(56), 56.

Cervellin, G., Comelli, I., & Lippi, G. (2017). Is Google Trends a reliable tool for digital epidemiology? Insights from different clinical settings. *Journal of epidemiology and global health, 7*(3), 185-189.

Chinazzi, M., Davis, J. T., Ajelli, M., Gioannini, C., Litvinova, M., Merler, S., . . . Sun, K. (2020). The effect of travel restrictions on the spread of the 2019 novel coronavirus (COVID-19) outbreak. *Science, 368*(6489), 395-400.

Colizza, V., Barrat, A., Barthelemy, M., Valleron, A.-J., & Vespignani, A. (2007). Modeling the worldwide spread of pandemic influenza: baseline case and containment interventions. *PLoS medicine, 4*(1).

Cook, S., Conrad, C., Fowlkes, A. L., & Mohebbi, M. H. (2011). Assessing Google flu trends performance in the United States during the 2009 influenza virus A (H1N1) pandemic. *PloS one, 6*(8).

Cowling, B. J., Ali, S. T., Ng, T. W., Tsang, T. K., Li, J. C., Fong, M. W., . . . Chiu, S. S. (2020). Impact assessment of non-pharmaceutical interventions against coronavirus disease 2019 and influenza in Hong Kong: an observational study. *The Lancet Public Health*.

Dandekar, R., & Barbastathis, G. (2020). Neural Network aided quarantine control model estimation of global Covid-19 spread. *arXiv preprint arXiv:2004.02752*.

Davies, N. G., Kucharski, A. J., Eggo, R. M., Gimma, A., Edmunds, W. J., Jombart, T., . . . Nightingale, E. S. (2020). Effects of non-pharmaceutical interventions on COVID-19 cases, deaths, and demand for hospital services in the UK: a modelling study. *The Lancet Public Health*.

de Arruda, G. F., Petri, G., & Moreno, Y. (2020). Social contagion models on hypergraphs. *Physical Review Research, 2*(2), 023032.

de Arruda, G. F., Petri, G., Rodrigues, F. A., & Moreno, Y. (2020). Impact of the distribution of recovery rates on disease spreading in complex networks. *Physical Review Research, 2*(1), 013046.

De Montjoye, Y.-A., Hidalgo, C. A., Verleysen, M., & Blondel, V. D. (2013). Unique in the crowd: The privacy bounds of human mobility. *Scientific reports, 3*, 1376.

Dietz, K. (1993). The estimation of the basic reproduction number for infectious diseases. *Statistical methods in medical research, 2*(1), 23-41.

Dolgin, E. (2010). Targeting hotspots of transmission promises to reduce malaria. In: Nature Publishing Group.

Dong, E., Du, H., & Gardner, L. (2020). An interactive web-based dashboard to track COVID-19 in real time. *The Lancet Infectious Diseases, 20*(5), 533-534.

Eames, K. T., & Keeling, M. J. (2003). Contact tracing and disease control. *Proceedings of the Royal Society of London. Series B: Biological Sciences, 270*(1533), 2565-2571.

Espana, G., Cavany, S., Oidtman, R. J., Barbera, C., Costello, A., Lerch, A., . . . Moore, S. M. (2020). Impacts of K-12 school reopening on the COVID-19 epidemic in Indiana, USA. *medRxiv*.

Eubank, S., Guclu, H., Kumar, V. A., Marathe, M. V., Srinivasan, A., Toroczkai, Z., & Wang, N. (2004). Modelling disease outbreaks in realistic urban social networks. *nature, 429*(6988), 180-184.

Eysenbach, G. (2009). Infodemiology and infoveillance: framework for an emerging set of public health informatics methods to analyze search, communication and publication behavior on the Internet. *Journal of medical Internet research, 11*(1), e11.

Ferretti, L., Wymant, C., Kendall, M., Zhao, L., Nurtay, A., Abeler-Dörner, L., . . . Fraser, C. (2020). Quantifying SARS-CoV-2 transmission suggests epidemic control with digital contact tracing. *Science, 368*(6491).

Finger, F., Genolet, T., Mari, L., de Magny, G. C., Manga, N. M., Rinaldo, A., & Bertuzzo, E. (2016). Mobile phone data highlights the role of mass gatherings in the spreading of cholera outbreaks. *Proceedings of the National Academy of Sciences, 113*(23), 6421-6426.

Flaxman, S., Mishra, S., Gandy, A., Unwin, H. J. T., Mellan, T. A., Coupland, H., . . . Eaton, J. W. (2020). Estimating the effects of non-pharmaceutical interventions on COVID-19 in Europe. *nature*, 1-5.

Fong, S. J., Li, G., Dey, N., Crespo, R. G., & Herrera-Viedma, E. (2020). Finding an accurate early forecasting model from small dataset: A case of 2019-ncov novel coronavirus outbreak. *arXiv preprint arXiv:2003.10776*.

General, U. S. (2019). The age of digital interdependence. In: Report of the UN Secretary-General's High-Level Panel on Digital Cooperation ….

General, U. S. (2020). Good communication saves lives. Retrieved from https://www.un.org/en/coronavirus/good-communication-saves-lives

Ginsberg, J., Mohebbi, M. H., Patel, R. S., Brammer, L., Smolinski, M. S., & Brilliant, L. (2009). Detecting influenza epidemics using search engine query data. *nature, 457*(7232), 1012-1014.

Glass, R. J., Glass, L. M., Beyeler, W. E., & Min, H. J. (2006). Targeted social distancing designs for pandemic influenza. *Emerging infectious diseases, 12*(11), 1671.

Gonzalez, M. C., Hidalgo, C. A., & Barabasi, A.-L. (2008). Understanding individual human mobility patterns. *nature, 453*(7196), 779-782.

Green, M. D., Freedman, D. M., & Gordis, L. (2000). Reference guide on epidemiology. *Reference Manual on Scientific Evidence, 2*, 638.

Greenwood, F., Howarth, C., Escudero Poole, D., Raymond, N. A., & Scarnecchia, D. P. (2017). The signal code: A human rights approach to information during crisis. *Harvard, MA*.

Hartono, P. (2020). Generating Similarity Map for COVID-19 Transmission Dynamics with Topological Autoencoder. *arXiv preprint arXiv:2004.01481*.

Hernán, M. A., & Robins, J. M. (2006). Estimating causal effects from epidemiological data. *Journal of Epidemiology & Community Health, 60*(7), 578-586.

Hu, Z., Ge, Q., Jin, L., & Xiong, M. (2020). Artificial intelligence forecasting of covid-19 in china. *arXiv preprint arXiv:2002.07112*.

Hu, Z., Ge, Q., Li, S., Boerwincle, E., Jin, L., & Xiong, M. (2020). Forecasting and evaluating intervention of Covid-19 in the World. *arXiv preprint arXiv:2003.09800*.

Hua, J., & Shaw, R. (2020). Corona virus (Covid-19)"infodemic" and emerging issues through a data lens: The case of china. *International journal of environmental research and public health, 17*(7), 2309.

Huang, C.-J., Chen, Y.-H., Ma, Y., & Kuo, P.-H. (2020). Multiple-Input Deep Convolutional Neural Network Model for COVID-19 Forecasting in China. *medRxiv*.

Imran, A., Posokhova, I., Qureshi, H. N., Masood, U., Riaz, S., Ali, K., . . . Nabeel, M. (2020). AI4COVID-19: AI enabled preliminary diagnosis for COVID-19 from cough samples via an app. *Informatics in Medicine Unlocked*, 100378.

Kermali, M., Khalsa, R. K., Pillai, K., Ismail, Z., & Harky, A. (2020). The role of biomarkers in diagnosis of COVID-19–A systematic review. *Life Sciences*, 117788.

Kuhn, C., Beck, M., & Strufe, T. (2020). Covid Notions: Towards Formal Definitions--and Documented Understanding--of Privacy Goals and Claimed Protection in Proximity-Tracing Services. *arXiv preprint arXiv:2004.07723*.

Kupferschmidt, K. (2020). Why do some COVID-19 patients infect many others, whereas most don't spread the virus at all? *Science*.

Lai, S., Ruktanonchai, N. W., Zhou, L., Prosper, O., Luo, W., Floyd, J. R., . . . Du, X. (2020). Effect of non-pharmaceutical interventions to contain COVID-19 in China.

Lampos, V., Moura, S., Yom-Tov, E., Cox, I. J., McKendry, R., & Edelstein, M. (2020). Tracking COVID-19 using online search. *arXiv preprint arXiv:2003.08086*.

Lazer, D., Kennedy, R., King, G., & Vespignani, A. (2014). The parable of Google Flu: traps in big data analysis. *Science, 343*(6176), 1203-1205.

Lazer, D. M., Baum, M. A., Benkler, Y., Berinsky, A. J., Greenhill, K. M., Menczer, F., . . . Rothschild, D. (2018). The science of fake news. *Science, 359*(6380), 1094-1096.

LeCun, Y., Bengio, Y., & Hinton, G. (2015). Deep learning. *nature, 521*(7553), 436-444.

Leung, G. M., & Leung, K. (2020). Crowdsourcing data to mitigate epidemics. *The Lancet Digital Health, 2*(4), e156-e157.

Liu, D., Clemente, L., Poirier, C., Ding, X., Chinazzi, M., Davis, J. T., . . . Santillana, M. (2020). A machine learning methodology for real-time forecasting of the 2019-2020 COVID-19 outbreak using Internet searches, news alerts, and estimates from mechanistic models. *arXiv preprint arXiv:2004.04019*.

Liu, Q.-H., Ajelli, M., Aleta, A., Merler, S., Moreno, Y., & Vespignani, A. (2018). Measurability of the epidemic reproduction number in data-driven contact networks. *Proceedings of the National Academy of Sciences, 115*(50), 12680-12685.

Lloyd-Smith, J. O., Schreiber, S. J., Kopp, P. E., & Getz, W. M. (2005). Superspreading and the effect of individual variation on disease emergence. *nature, 438*(7066), 355-359.

Lu, F. S., Hattab, M. W., Clemente, C. L., Biggerstaff, M., & Santillana, M. (2019). Improved state-level influenza nowcasting in the United States leveraging Internet-based data and network approaches. *Nature communications, 10*(1), 1-10.

Luengo-Oroz, M., Pham, K. H., Bullock, J., Kirkpatrick, R., Luccioni, A., Rubel, S., . . . Nguyen, T. (2020). Artificial intelligence cooperation to support the global response to COVID-19. *Nature Machine Intelligence*, 1-3.

Lynch, C., & Roper, C. (2011). The transit phase of migration: circulation of malaria and its multidrug-resistant forms in Africa. *PLoS medicine, 8*(5).

Malone, T. W. (2004). *The future of work*: Audio-Tech Business Book Summaries, Incorporated.

Martín-Calvo, D., Aleta, A., Pentland, A., Moreno, Y., & Moro, E. (2020). *Effectiveness of social distancing strategies for protecting a community from a pandemic with a data driven contact network based on census and real-world mobility data*. Retrieved from

Mejova, Y., & Kalimeri, K. (2020). Advertisers jump on coronavirus bandwagon: Politics, news, and business. *arXiv preprint arXiv:2003.00923*.

Mello, M. M., & Wang, C. J. (2020). Ethics and governance for digital disease surveillance. *Science*.

Meloni, S., Perra, N., Arenas, A., Gómez, S., Moreno, Y., & Vespignani, A. (2011). Modeling human mobility responses to the large-scale spreading of infectious diseases. *Scientific reports, 1*, 62.

Menni, C., Valdes, A., Freydin, M. B., Ganesh, S., Moustafa, J. E.-S., Visconti, A., . . . Falchi, M. (2020). Loss of smell and taste in combination with other symptoms is a strong predictor of COVID-19 infection. *medRxiv*.

Meyers, L. A., Newman, M., & Pourbohloul, B. (2006). Predicting epidemics on directed contact networks. *Journal of theoretical biology, 240*(3), 400-418.

Miritello, G., Moro, E., & Lara, R. (2011). Dynamical strength of social ties in information spreading. *Physical Review E, 83*(4), 045102.

Mizumoto, K., Kagaya, K., Zarebski, A., & Chowell, G. (2020). Estimating the asymptomatic proportion of coronavirus disease 2019 (COVID-19) cases on board the Diamond Princess cruise ship, Yokohama, Japan, 2020. *Eurosurveillance, 25*(10), 2000180.

Morales, A. J., Borondo, J., Losada, J. C., & Benito, R. M. (2014). Efficiency of human activity on information spreading on Twitter. *Social Networks, 39*, 1-11.

Moreno, Y., Pastor-Satorras, R., & Vespignani, A. (2002). Epidemic outbreaks in complex heterogeneous networks. *The European Physical Journal B-Condensed Matter and Complex Systems, 26*(4), 521-529.

Nekovee, M., Moreno, Y., Bianconi, G., & Marsili, M. (2007). Theory of rumour spreading in complex social networks. *Physica A: Statistical Mechanics and its Applications, 374*(1), 457-470.

Newman, M. E. (2002). Spread of epidemic disease on networks. *Physical Review E, 66*(1), 016128.

Newman, M. E., Barabási, A.-L. E., & Watts, D. J. (2006). *The structure and dynamics of networks*: Princeton university press.

Nicholas, G. D., Petra, K., Yang, L., Kiesha, P., Mark, J., Rosalind, M., & group, C. C.-w. (2020). Age-dependent Effects in the Transmission and Control of COVID-19 Epidemics. *Nature medicine*.

Norman, J., Bar-Yam, Y., & Taleb, N. N. (2020). Systemic Risk of Pandemic via Novel Pathogens—Coronavirus: A Note. *New England Complex Systems Institute (January 26, 2020)*.

Oliver, N., Lepri, B., Sterly, H., Lambiotte, R., Delataille, S., De Nadai, M., . . . Cattuto, C. (2020). Mobile phone data for informing public health actions across the COVID-19 pandemic life cycle. In: American Association for the Advancement of Science.

Olson, D. R., Konty, K. J., Paladini, M., Viboud, C., & Simonsen, L. (2013). Reassessing Google Flu Trends data for detection of seasonal and pandemic influenza: a comparative epidemiological study at three geographic scales. *PLoS computational biology, 9*(10).

Pastor-Escuredo, D. (2020). Ethics in the digital era. *arXiv preprint arXiv:2003.06530*.

Pastor-Escuredo, D., & Frias-Martinez, E. (2020). Flow descriptors of human mobility networks. *arXiv preprint arXiv:2003.07279*.
Pastor-Escuredo, D., & Tarazona, C. (2020). Characterizing information leaders in Twitter during COVID-19 crisis. *arXiv preprint arXiv:2005.07266*.
Pastor-Satorras, R., & Vespignani, A. (2001a). Epidemic dynamics and endemic states in complex networks. *Physical Review E, 63*(6), 066117.
Pastor-Satorras, R., & Vespignani, A. (2001b). Epidemic spreading in scale-free networks. *Physical review letters, 86*(14), 3200.
Pastor-Satorras, R., & Vespignani, A. (2002). Epidemic dynamics in finite size scale-free networks. *Physical Review E, 65*(3), 035108.
Perrota, D. (2018). Can Mobile Phone Traces Help Shed Light on the Spread of Zika in Colombia? Retrieved from https://www.unglobalpulse.org/2018/04/can-mobile-phone-traces-help-shed-light-on-the-spread-of-zika-in-colombia/
Popat, K., Mukherjee, S., Yates, A., & Weikum, G. (2018). DeClarE: Debunking fake news and false claims using evidence-aware deep learning. *arXiv preprint arXiv:1809.06416*.
Prothero, R. M. (1977). Disease and mobility: a neglected factor in epidemiology. *International journal of epidemiology, 6*(3), 259-267.
Pulse, U. G. (2012). Big data for development: Challenges & opportunities. *Naciones Unidas, Nueva York, mayo*.
Pulse, U. G. (2015). Mapping the risk-utility landscape: mobile data for sustainable development and humanitarian action. *Global Pulse Project Series no18*.
Rajarethinam, J., Ong, J., Lim, S.-H., Tay, Y.-H., Bounliphone, W., Chong, C.-S., . . . Ng, L.-C. (2019). Using human movement data to identify potential areas of Zika transmission: case study of the largest Zika cluster in Singapore. *International journal of environmental research and public health, 16*(5), 808.
Riley, S. (2007). Large-scale spatial-transmission models of infectious disease. *Science, 316*(5829), 1298-1301.
Ruchansky, N., Seo, S., & Liu, Y. (2017). *Csi: A hybrid deep model for fake news detection.* Paper presented at the Proceedings of the 2017 ACM on Conference on Information and Knowledge Management.
Salathé, M. (2018). Digital epidemiology: what is it, and where is it going? *Life sciences, society and policy, 14*(1), 1.
Salathe, M., Bengtsson, L., Bodnar, T. J., Brewer, D. D., Brownstein, J. S., Buckee, C., . . . Mabry, P. L. (2012). Digital epidemiology. *PLoS computational biology, 8*(7).
Sanz, J., Xia, C.-Y., Meloni, S., & Moreno, Y. (2014). Dynamics of interacting diseases. *Physical Review X, 4*(4), 041005.
Shi, Y., Wang, Y., Shao, C., Huang, J., Gan, J., Huang, X., . . . Melino, G. (2020). COVID-19 infection: the perspectives on immune responses. In: Nature Publishing Group.
Shu, K., Sliva, A., Wang, S., Tang, J., & Liu, H. (2017). Fake news detection on social media: A data mining perspective. *ACM SIGKDD Explorations Newsletter, 19*(1), 22-36.
Simini, F., González, M. C., Maritan, A., & Barabási, A.-L. (2012). A universal model for mobility and migration patterns. *nature, 484*(7392), 96.
Singh, L., Bansal, S., Bode, L., Budak, C., Chi, G., Kawintiranon, K., . . . Wang, Y. (2020). A first look at COVID-19 information and misinformation sharing on Twitter. *arXiv preprint arXiv:2003.13907*.

Singhania, S., Fernandez, N., & Rao, S. (2017). *3han: A deep neural network for fake news detection.* Paper presented at the International Conference on Neural Information Processing.

Stoddard, S. T., Morrison, A. C., Vazquez-Prokopec, G. M., Soldan, V. P., Kochel, T. J., Kitron, U., . . . Scott, T. W. (2009). The role of human movement in the transmission of vector-borne pathogens. *PLoS neglected tropical diseases, 3*(7).

Sun, K., Chen, J., & Viboud, C. (2020). Early epidemiological analysis of the coronavirus disease 2019 outbreak based on crowdsourced data: a population-level observational study. *The Lancet Digital Health*.

Taleb, N. N. (2019). The Statistical Consequences of Fat Tails. In: STEM Publishing.

Taleb, N. N., Bar-Yam, Y., Douady, R., Norman, J., & Read, R. (2014). The precautionary principle: fragility and black swans from policy actions. *NYU Extreme Risk Initiative Working Paper*, 1-24.

Taleb, N. N., Read, R., Douady, R., Norman, J., & Bar-Yam, Y. (2014). The precautionary principle (with application to the genetic modification of organisms). *arXiv preprint arXiv:1410.5787*.

Tatem, A. J., Huang, Z., Narib, C., Kumar, U., Kandula, D., Pindolia, D. K., . . . Uusiku, P. (2014). Integrating rapid risk mapping and mobile phone call record data for strategic malaria elimination planning. *Malaria journal, 13*(1), 52.

Tizzoni, M., Bajardi, P., Decuyper, A., King, G. K. K., Schneider, C. M., Blondel, V., . . . Colizza, V. (2014). On the use of human mobility proxies for modeling epidemics. *PLoS computational biology, 10*(7).

UNICEF. (2020). Magic Box COVID-19 report. Retrieved from https://www.unicef.org/innovation/magicbox/covid

Vaezi, A., & Javanmard, S. H. (2020). Infodemic and risk communication in the era of CoV-19. *Advanced Biomedical Research, 9*.

Vespignani, A., Tian, H., Dye, C., Lloyd-Smith, J. O., Eggo, R. M., Shrestha, M., . . . Wu, J. (2020). Modelling COVID-19. *Nature Reviews Physics*, 1-3.

Vinuesa, R., Theodorou, A., Battaglini, M., & Dignum, V. (2020). A socio-technical framework for digital contact tracing. *arXiv preprint arXiv:2005.08370*.

Wesolowski, A., Buckee, C. O., Engø-Monsen, K., & Metcalf, C. J. E. (2016). Connecting mobility to infectious diseases: the promise and limits of mobile phone data. *The Journal of infectious diseases, 214*(suppl_4), S414-S420.

Wesolowski, A., Eagle, N., Tatem, A. J., Smith, D. L., Noor, A. M., Snow, R. W., & Buckee, C. O. (2012). Quantifying the impact of human mobility on malaria. *Science, 338*(6104), 267-270.

Wesolowski, A., Qureshi, T., Boni, M. F., Sundsøy, P. R., Johansson, M. A., Rasheed, S. B., . . . Buckee, C. O. (2015). Impact of human mobility on the emergence of dengue epidemics in Pakistan. *Proceedings of the National Academy of Sciences, 112*(38), 11887-11892.

Wesolowski, A., Stresman, G., Eagle, N., Stevenson, J., Owaga, C., Marube, E., . . . Buckee, C. O. (2014). Quantifying travel behavior for infectious disease research: a comparison of data from surveys and mobile phones. *Scientific reports, 4*, 5678.

WHO. (2020). Infodemic management - Infodemiology;. Retrieved from https://www.who.int/teams/risk-communication/infodemic-management.

Yang, S., Santillana, M., & Kou, S. C. (2015). Accurate estimation of influenza epidemics using Google search data via ARGO. *Proceedings of the National Academy of Sciences, 112*(47), 14473-14478.

Zarocostas, J. (2020). How to fight an infodemic. *The Lancet, 395*(10225), 676.

Zhang, J., Litvinova, M., Liang, Y., Wang, Y., Wang, W., Zhao, S., . . . Vespignani, A. (2020). Changes in contact patterns shape the dynamics of the COVID-19 outbreak in China. *Science*.

Zou, H., & Hastie, T. (2005). Regularization and variable selection via the elastic net. *Journal of the royal statistical society: series B (statistical methodology), 67*(2), 301-320.

Zufiria, P. J., Pastor-Escuredo, D., Úbeda-Medina, L., Hernandez-Medina, M. A., Barriales-Valbuena, I., Morales, A. J., . . . Quinn, J. (2018). Identifying seasonal mobility profiles from anonymized and aggregated mobile phone data. Application in food security. *PloS one, 13*(4), e0195714.